\documentclass[%
 reprint,
superscriptaddress,
 amsmath,amssymb,
 aps,
 prx
]{revtex4-1}

\usepackage{graphicx}
\usepackage{dcolumn}
\usepackage{bm}

\begin{document}

\preprint{APS/123-QED}

\title{Time delays from one-photon transitions in the continuum}

\author{Jaco Fuchs}
\email{jafuchs@phys.ethz.ch}
\affiliation{Department of Physics, Eidgen\"ossische Technische Hochschule Zürich, Zürich, Switzerland}

\author{Nicolas Douguet}%
\affiliation{Department of Physics, University of Central Florida, Orlando, Florida, USA}

\author{Stefan Donsa}
\affiliation{Institute of Theoretical Physics, Vienna University of Technology, Vienna, Austria, EU}

\author{Fernando Mart\'{\i}n}
\affiliation{Departamento de Quimica Modulo 13, Universidad Autonoma de Madrid, 28049 Madrid, Spain, EU}
\affiliation{Condensed Matter Physics Center (IFIMAC), Universidad Autonoma de Madrid, 28049 Madrid, Spain, EU}
\affiliation{Instituto Madrileno de Estudios Avanzados en Nanociencia (IMDEA-Nano), 28049 Madrid, Spain, EU}

\author{Joachim Burgd\"orfer}
\affiliation{Institute of Theoretical Physics, Vienna University of Technology, Vienna, Austria, EU}

\author{Luca Argenti}
\affiliation{Department of Physics, University of Central Florida, Orlando, Florida, USA}
\affiliation{CREOL, University of Central Florida, Orlando, Florida 32186, USA}

\author{Laura Cattaneo}
\affiliation{Department of Physics, Eidgen\"ossische Technische Hochschule Zürich, Zürich, Switzerland}

\author{Ursula Keller}
\affiliation{Department of Physics, Eidgen\"ossische Technische Hochschule Zürich, Zürich, Switzerland}


\date{\today}

\begin{abstract}
Attosecond photoionisation time delays reveal information about the potential energy landscape an outgoing electron wavepacket probes upon ionisation. In this study we experimentally quantify, for the first time, the dependence of the time delay on the angular momentum of the liberated photoelectrons. For this purpose, electron quantum-path interference spectra have been resolved in energy and angle using a two-color attosecond pump-probe photoionisation experiment in helium. A fitting procedure of the angle-dependent interference pattern allows us to disentangle the relative phase of all four quantum pathways that are known to contribute to the final photoelectron signal. In particular, we resolve the dependence on the angular momentum of the delay of one-photon transitions between continuum states, which is an essential and universal contribution to the total photoionization delay observed in attosecond pump-probe measurements. For such continuum-continuum transitions, we measure a delay between outgoing $s$- and $d$-electrons as large as 12 as close to the ionisation threshold in helium. Both single-active-electron and first-principles ab initio simulations confirm this observation for helium and hydrogen, demonstrating the universality of the observed delays.
\end{abstract}

\maketitle


\section{Introduction}

Free electrons cannot exchange photons with a light pulse. Unbound electrons, however, which are subject to an external potential, can absorb (inverse Bremsstrahlung) or emit (stimulated Bremsstrahlung) quanta of the radiation field. In the presence of an attractive Coulomb potential of a nearby ion the absorption and emission of a single photon promote dipole transitions that change the quantum state. These transitions involve bound as well as continuum states, giving rise to various types of radiative processes such as excitation (bound to bound), ionisation (bound to continuum, bc), recombination (continuum to bound),  and continuum-continuum (cc) transitions.

Recent progress in attosecond science has given direct access to timing information in photon-atom interaction on the attosecond scale. In particular, single photon ionisation and the corresponding Eisenbud-Wigner-Smith (EWS) delay  \cite{Wigner1955, Smith1960, Dahlstrom2013, Pazourek2015b} attracted lots of attention. Briefly, due to the propagation across the potential-energy landscape the excited photoelectron wave packet aquires an energy-dependent phase which results in a measurable group delay, referred to as a photoionisation time delay. Relative delays between wave packets from different species \cite{Sabbar2015, Huppert2016, Sabbar2017a}, ionisation channels \cite{Cavalieri2007, Schultze2010, Sansone2010, Ossiander2016, Vos2018, Cattaneo2018}, and emission angles \cite{Heuser2016, Cirelli2018}, have been measured to very high accuracy and serve as benchmarks for time-dependent quantum mechanical simulations in atoms \cite{Schultze2010, Sabbar2015, Ossiander2016, Heuser2016, Cirelli2018, Sabbar2017a}, molecules \cite{Sansone2010, Huppert2016, Vos2018, Cattaneo2018}, and solids \cite{Miaja-Avila2006, Cavalieri2007}. 

To date these attosecond measurement techniques are based on the delay between two coherent laser pulses which are typically in the extreme ultraviolet (XUV) and the infrared (IR). Thus the time delays could only be measured between ionisation pathways involving at least two photons. In particular, if the first photoabsorption event is a bound-continuum (bc) transition a second transition in the continuum is required to access temporal information in state-of-the-art experiments. Thus, in addition to the EWS delay \cite{Wigner1955, Smith1960}, the experimentally observed delays contain two more contributions. The first one originates from the spectral phase of the ionising attosecond pulse train (APT) \cite{Mairesse2003, Cattaneo2016a} and cancels out when comparing different species or channels. The second contribution originates from the cc-transitions mediated by the probing IR laser pulse \cite{Dahlstrom2012,Pazourek2015b}. Although it is well known \cite{Klunder2011, Dahlstrom2013, Ivanov2017} that the cc-contribution to the photoionisation time delay can be comparable or even larger than the EWS delay for single photon ionisation, it has surprisingly drawn much less attention. Moreover, since experimentally disentangling the contributions has not been possible so far, time delays of one-photon ionization were only accessible when referencing to theoretical calculations \cite{Klunder2011, Nagele2011}.

\begin{figure}[t]
\includegraphics[width=0.48\textwidth]{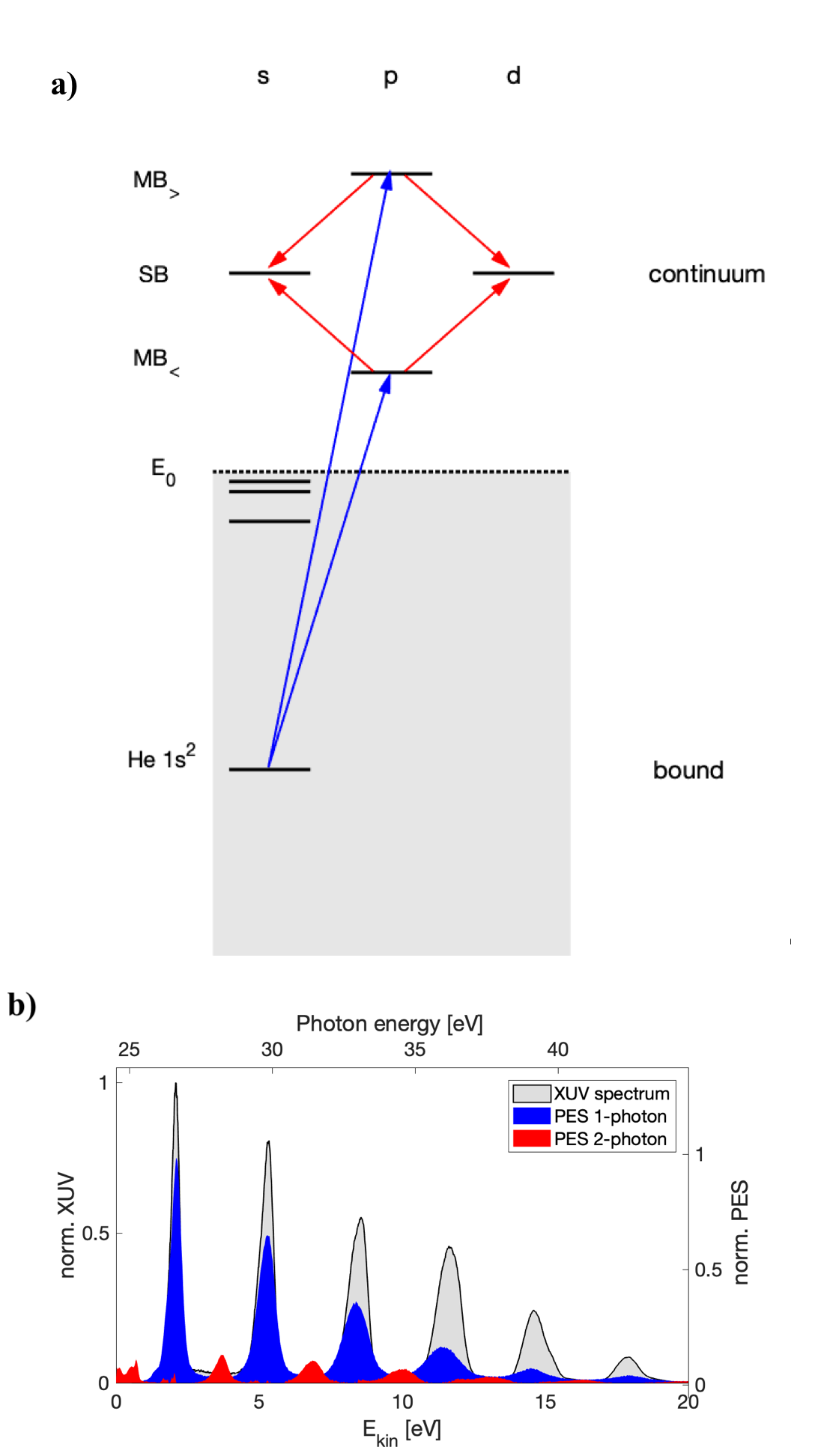}
\caption{\label{fig1} (a) Quantum pathways leading to the same sideband. The XUV (blue) mediates bound to continuum transitions (ionisation), the IR (red) mediates transitions within the continuum. All quantum pathways interfere.  (b) Experimental XUV and photoelectron energy  spectra (PES).  The one-photon PES is obtained by an XUV only measurement. The two-photon PES is obtained by subtracting the one-photon spectrum from the time-integrated RABBITT spectrum.}
\end{figure}

Recently, experimental evidence of a strong effect of the IR-induced cc-transitions on the angular dependence of the total photoemission delays has been reported \cite{Heuser2016, Cirelli2018} stimulating several independent investigations on the origin of this effect \cite{Watzel2015,  Ivanov2017, Hockett2017a, Busto2018, Bray2018,Zhao2018}. In this work, we present a new method that allows us to unravel the delay between electron wave packets from different one photon transitions in the continuum, purely from experimental data, and independently of the Wigner and XUV contributions.  We obtain for first time access to the angular momentum dependence of the EWS delay for cc-transitions. The method is based on an algorithm developed to analyze angularly resolved RABBITT (Reconstruction of Attosecond Beating By Interference of Two-photon Transitions \cite{Paul2001}) spectra. We find an ubiquitous positive and energy-dependent time delay, as large as 12 as, between $s$- and $d$-wave photoelectrons produced by the additional IR-photon exchange that follows photoionisation of atomic helium by an XUV attosecond pulse train. This result is the first demonstration of a direct measurement of the EWS delay arising from one-photon transitions within the continuum. Using two independent computational methods to solve the time-dependent Schrödinger equation (TDSE), one based on the single-active-electron approximation and the other being a first-principles \textit{ab initio} approach, we obtain excellent agreement with the experimentally retrieved ionization time delays. These findings confirm that, in helium, at energies close to the first ionization threshold, the delay associated to radiative transitions in the continuum is dominated by the electron angular momentum and radial momentum distribution, whereas electronic correlation plays no significant role. 

The following section develops the theoretical framework needed to interpret the experiment and the simulations. The experimental and theoretical analysis are described in sections 3 and 4, respectively. The main results will be examined in section 5 and the last section we will offer our conclusions.

\section{Theoretical framework}

In the photoionisation of helium from its $1s^2$ ground state by absorption of a single XUV photon, the total angular momentum of the combined atom-photon system has to be conserved. In the energy range examined in this work, the residual He$^+$ ion remains in its lowest $^2S$ (1s) state. The photon angular momentum, therefore, is entirely transferred to the ejected electron, which is emitted as a p-wave (angular momentum l=1). This liberated electron can subsequently absorb (or emit) an additional IR photon through a cc-transition, which transfers to the photoelectron additional angular momentum resulting in either an $s$-wave (l=0) or $d$-wave (l=2). For parallel polarized XUV and IR pulses the magnetic quantum number remains zero.

In the RABBITT technique \cite{Paul2001, Muller2002}, an XUV pulse train  consisting of odd harmonics of the fundamental IR laser frequency $\omega$ is used to ionise the target, leading to single-photon peaks (mainbands) in the photoelectron spectrum separated by twice the laser photon energy $2\hbar\omega$. A weak replica of the fundamental IR with frequency $\omega$ then triggers cc-transitions from the mainband to sidebands with kinetic  energies lying between the mainbands. For ionization from an s-shell, four main quantum pathways contribute to each sideband, namely, the transitions $s\rightarrow p \rightarrow s$ and $s \rightarrow p \rightarrow d$, for both the absorption and stimulated emission of the IR photon, as illustrated schematically in Figure \ref{fig1}a. At low IR intensities (less than few GW/cm$^2$ at 800 nm) pathways to the sidebands that involve the exchange of more than one IR photon give a negligible contribution and hence states with angular momentum higher than 2 are not populated. Due to the interference of absorption and emission pathways, the sideband signal oscillates as a function of the delay $\tau$ between the IR pulse and the XUV pulse train. In Fig. \ref{fig1}b we show the spectrum of the XUV pulse train and the corresponding photoelectron energy spectrum (PES) of one-photon and two-photon pathways.  

In the weak-field regime, the $s$ (l=0) and $d$ (l=2) photoionisation amplitudes at the sideband with energy $E_f$ can be expressed, within lowest order of perturbation theory, by the well-known two-photon-transition formula \cite{Jimenez-Galan2016} 
\begin{equation}\label{eq1}
\begin{split}
A^{(2)}_l(E_f)=&-i\int d\Omega M^{(2)}_{E_f,l}(\Omega)E_{XUV}(\Omega)E_{IR}([E_f-I_p]/\hbar-\Omega)\\
=&\hspace{10pt}A^{(2+)}_l+A^{(2-)}_l
\end{split}
\end{equation}
where $M_{E_f,l}^{(2)}$ is the two-photon matrix element, $I_p$ is the ionisation potential and $E_{XUV}$ and $E_{IR}$ are the Fourier transforms of the XUV and IR electric field, respectively.

The frequency (energy) integral in Eq. \eqref{eq1} can be split into an interval with $0 < \Omega < (E_f-I_p)/ \hbar$, $A_l^{(2+)}$, which corresponds to pathways with absorption of an IR photon and an integral with $\Omega \geq (E_f-I_p)/\hbar$, $A_l^{(2-)}$, for the pathways with stimulated emission of an IR photon.  Following \cite{Klunder2011}, for a narrow-band IR spectrum with frequency $\omega$ far from resonances, the phase of the two-photon matrix element (Eq. \eqref{eq1}) can be decomposed into three additive contributions
\begin{equation}\label{eq2}
A_l^{(2\pm)}=|A_l^{(2\pm)}|e^{i(\varphi_l^{cc\pm}+\varphi_{\lessgtr}^{bc}\pm\omega\tau)} 
\end{equation}
with $\varphi_l^{cc\pm}$ the phase of the cc-transition with final angular momentum $l$, $\varphi_\lessgtr^{bc}$ the phase of the one-photon bc-transition to the lower ($<$) or upper ($>$) main band and the phase $\pm \omega \tau $ due to the pump-probe delay $\tau$, which leads to oscillations in the interference pattern. The resulting ionisation probability at the sideband is 
\begin{equation}\label{eq3}
\begin{split}
I(\vartheta,\varphi,\tau)&=|(A^{(2+)}_s+A^{(2-)}_s)Y^0_0(\vartheta,\varphi)\\
&\hspace{0.5cm}+(A^{(2+)}_d+A^{(2-)}_d)Y^0_2(\vartheta,\varphi)|^2\\
&=\sum_{n=0}^4\beta_n(\tau)P_n[\cos(\theta)]
\end{split}
\end{equation}
where $\theta$ is the angle between the common laser polarisation axis of the XUV and IR electric field, and the direction of the outgoing electron. The series expansion in Legendre polynomials $P_n$ extends up to fourth-order \cite{Laurent2012}. The coefficients $\beta_n (\tau)$, which quantify the photoemission anisotropy, have the following expressions
\begin{align}
\begin{split}\label{eq4}
\beta_0=&|\mathcal{A}_{s}^{(2+)}|^2+|\mathcal{A}_{s}^{(2-)}|^2+|\mathcal{A}_{d}^{(2+)}|^2+|\mathcal{A}_{d}^{(2-)}|^2\\
+&2|\mathcal{A}_{s}^{(2+)}||\mathcal{A}_{s}^{(2-)}|\cos(2\omega\tau+\varphi_{s}^{(2+)}-\varphi_{s}^{(2-)})\\
+&2|\mathcal{A}_{d}^{(2+)}||\mathcal{A}_{d}^{(2-)}|\cos(2\omega\tau+\varphi_{d}^{(2+)}-\varphi_{d}^{(2-)}),
\end{split}\\[10pt]
\begin{split}\label{eq5}
\beta_2=&\frac{10}{7}[|\mathcal{A}_{d}^{(2+)}|^2+|\mathcal{A}_{d}^{(2-)}|^2\\
&\hspace{0.0cm}+2|\mathcal{A}_{d}^{(2+)}||\mathcal{A}_{d}^{(2-)}|\cos(2\omega_\tau+\varphi_{d}^{(2+)}-\varphi_{d}^{(2-)})]\\
+&2\sqrt{5}[|\mathcal{A}_{s}^{(2+)}||\mathcal{A}_{d}^{(2+)}|\cos(\varphi_{s}^{(2+)}-\varphi_{d}^{(2+)})\\
&\hspace{0.3cm}+|\mathcal{A}_{s}^{(2-)}||\mathcal{A}_{d}^{(2-)}|\cos(\varphi_{s}^{(2-)}-\varphi_{d}^{(2-)})\\
&\hspace{0.3cm}+|\mathcal{A}_{s}^{(2+)}||\mathcal{A}_{d}^{(2-)}|\cos(2\omega\tau+\varphi_{s}^{(2+)}-\varphi_{d}^{(2-)})\\
&\hspace{0.3cm}+|\mathcal{A}_{s}^{(2-)}||\mathcal{A}_{d}^{(2+)}|\cos(2\omega\tau+\varphi_{d}^{(2+)}-\varphi_{s}^{(2-)})],
\end{split}\\[10pt]
\begin{split}\label{eq6}
\beta_4=&\frac{18}{7}[|\mathcal{A}_{d}^{(2+)}|^2+|\mathcal{A}_{d}^{(2-)}|^2\\
+&2|\mathcal{A}_{d}^{(2+)}||\mathcal{A}_{d}^{(2-)}|\cos(2\omega\tau+\varphi_{d}^{(2+)}-\varphi_{d}^{(2-)})].
\end{split}
\end{align}
Since s- and d- waves have the same (even) parity, the odd anisotropy parameters $\beta_1$ and $\beta_3$ are identically zero. Here
\begin{equation}\label{eq7}
\varphi_{s,d}^{(2\pm)}=\varphi_{s,d}^{cc\pm}+\varphi^{bc}_\lessgtr
\end{equation}
contains both the phase $\varphi_{s,d}^{cc\pm}$ of the cc-transition and the phase $\varphi_\lessgtr^{bc}$  associated with the preceding ionisation. The latter one contains the phase of the ionizing XUV pulse and the atomic phase $\delta_l^{bc}$ for the half-scattering process of the outgoing electron wavepacket at the atomic potential. Its spectral derivative $d\delta_l^{bc} (E)/dE$ gives the EWS delay for single photon ionization. Analogously  $d\varphi_{s,d}^{cc\pm}(E)/dE$ gives the delay of the cc-transition often referred to as cc-delay or Coulomb laser coupling (CLC) delay \cite{Pazourek2015b}. However, since the IR-driven cc-transition occurs primarily at large distances from the atomic core \cite{Klunder2011}, the accumulated phase, unlike for one-photon ionization does not account for the full half-scattering phase but only for the propagation in the long-range tail of the atomic potential. Therefore, the influence of the centrifugal potential $L^2/2r^2 $ on the cc-phase was previously neglected \cite{Klunder2011}.

\section{Experimental results}
\begin{figure}
\includegraphics[width=0.48\textwidth]{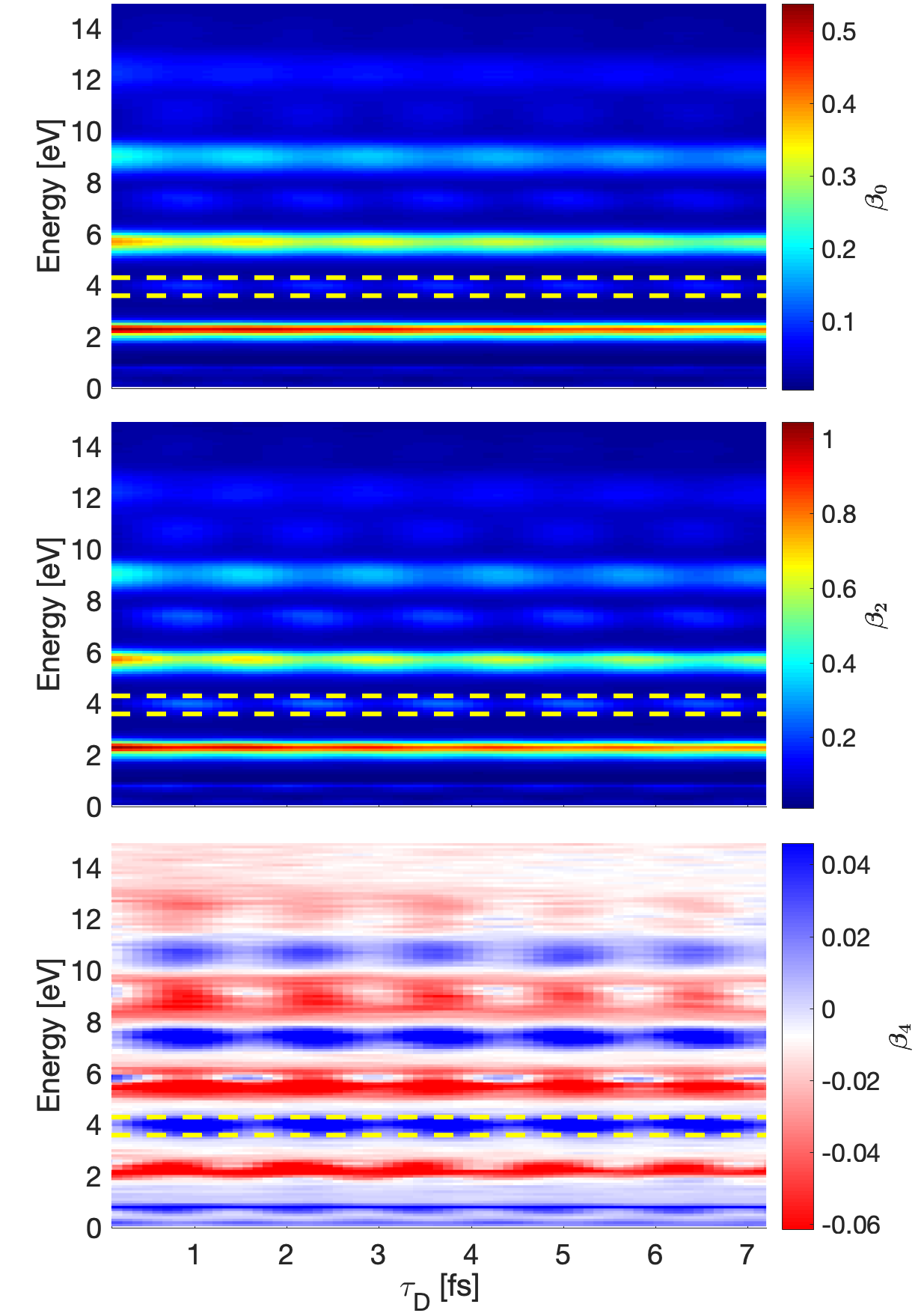}
\caption{\label{fig2} Experimental anisotropy parameters of the time resolved photoelectron angular distribution, $\beta_0$ (top), $\beta_2$ (centre) and $\beta_4$ (bottom). The yellow lines indicate the integration range of sideband 18. A positive delay indicates that the IR pulse is delayed with respect to the XUV, the zero delay is chosen arbitrarily.}
\end{figure}
In order to measure the time and angle-dependent ionisation probability (Eq. \eqref{eq3}), we use a COLTRIMS detector \cite{Dorner2000} in combination with an XUV-IR pump-probe setup. An amplified Ti-Sapphire laser, with a repetition rate of 10 kHz, generates a 790 nm IR pulse of 29 fs FWHM duration and 0.7 mJ total energy. The pulse is split into an intense (80\%) and a weaker (20\%) component. The stronger IR beam is focused into an argon gas cell where the XUV harmonics 13 to 25, corresponding to an energy range form 20 to 40 eV, are created by high harmonic generation \citep{Lewenstein1994} (see figure 1b). The remaining IR pulse passes through a delay stage and is recombined with the XUV beam, before being focused into a cold helium jet. With the COLTRIMS detector we measure the three-dimensional momentum of both photoelectrons and ions with a 4$\pi$ solid angle detection capability. With coincidence-selection (time-of-flight filtering of the helium ion and momentum conservation condition) we can discriminate against electrons from other reactions. The delay between the two pulses is controlled via a piezo driven delay stage in combination with an active interferometric stabilisation. For details on the experimental setup the reader is referred to \cite{Sabbar2014}.

 To guarantee a uniform detection capability over all emission angles and energies, we calibrate the detector efficiency using a helium XUV-only measurement, where the differential cross section is known exactly. For the time-resolved measurements, an IR intensity of $3\cdot 10^{11}$W/cm$2$ is used at the interaction region. The angle and time-dependent photoelectron spectra are recorded for 40 delay steps. The resulting angular distributions are then projected on the Legendre-polynomials [Eq.(3)] to retrieve the anisotropy parameters of the distribution, which are shown in Figure \ref{fig2}. The sideband signal is integrated over 0.5 eV, as indicated for sideband 18 by the yellow lines.

\begin{figure}
\includegraphics[width=0.48\textwidth]{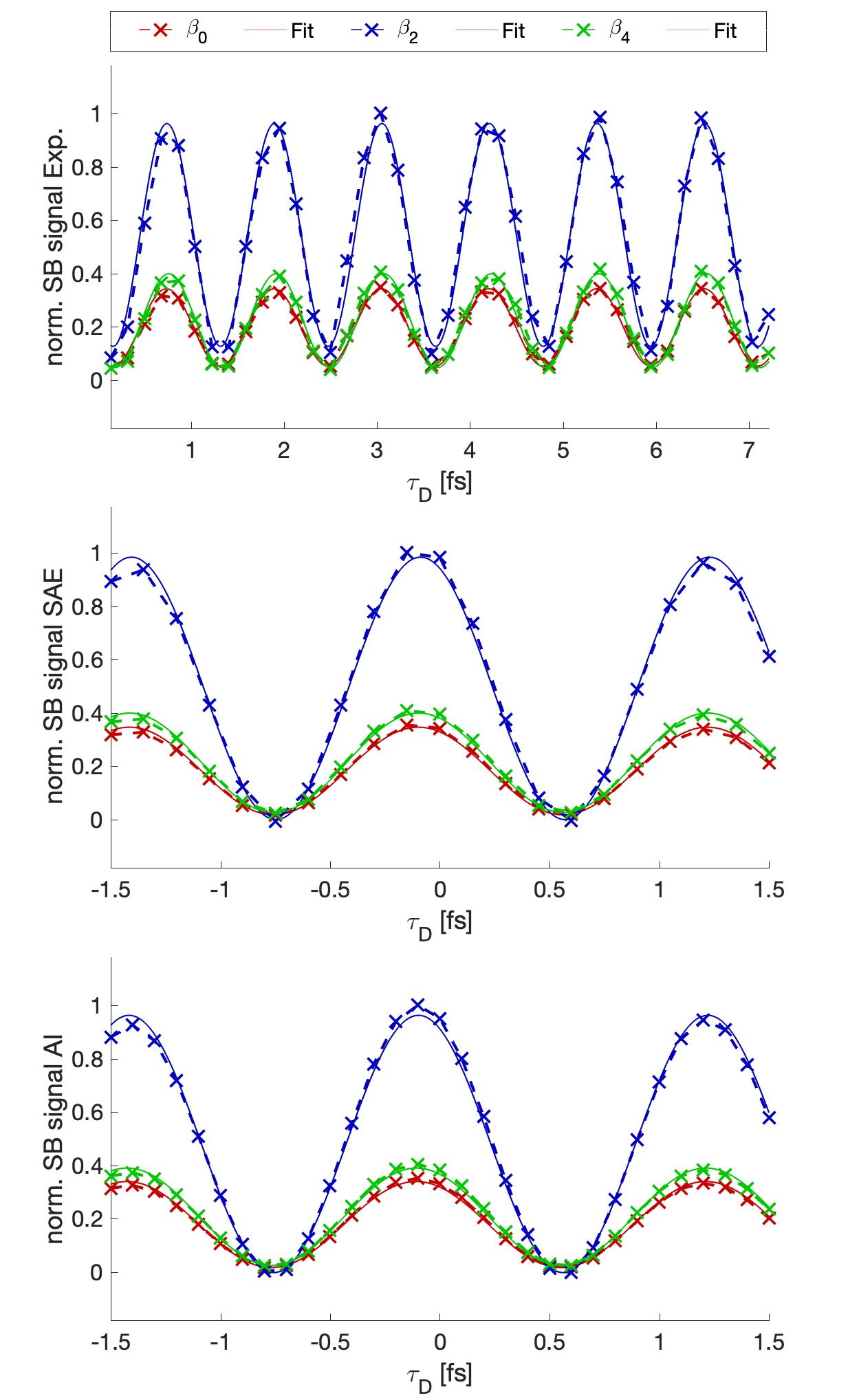}
\caption{\label{fig3} Simultaneous fit of the sideband anisotropy parameters. The time resolved sideband signal and the simultaneous fit of the anisotropy parameters $\beta_0$, $\beta_2$ and $\beta_4$  are shown for sideband 18 for the experimental data (top), the SAE calculations (middle) and the full ab initio (bottom), respectively.}
\end{figure}

Figure \ref{fig3} shows the anisotropy parameters, for each sideband, as a function of the time delay. Each of the three beta parameters oscillates at twice the IR frequency,  $\beta_n=a+b  \cos(2\omega\tau-\varphi)$, with offset $a$, amplitude $b$, and phase $\varphi$, which are directly related to the parameters in equations \eqref{eq4} – \eqref{eq6} and can all be unambiguously extracted from the measurement.

The system of equations \eqref{eq4}–\eqref{eq6} has in total 4 unknown amplitudes and 4 unknown phases. The phases appear in differences only, and are, thus, only determined up to an overall constant, allowing us to set one of the phases to zero without loss of generality. The remaining variables can then be simultaneously fitted to the system of equations using a least square minimization routine based on the Levenberg-Marquardt-algorithm \cite{More2006}. Figure \ref{fig3} shows the fit to the experimental data for sideband 18. The convergence of the fit to the correct set of parameters has been checked by performing the same fitting procedure on sets of simulated data. Making use of both the angle-dependent phase and amplitude of the RABBITT interference pattern we can thus determine the amplitudes and relative phases of all four quantum paths contributing to any given sideband. In particular, the relative phase between the two pathways which lead to different angular momenta, is for the absorption
\begin{equation}\label{eq8}
\begin{split}
\varphi_{s}^{(2+)}-\varphi_{d}^{(2+)}&=\varphi_{s}^{cc+}+\varphi^{bc}_{(<)}-\varphi_{d}^{cc+}-\varphi^{bc}_{(<)}\\
&=\varphi_{s}^{cc+}-\varphi_{d}^{cc+},
\end{split}
\end{equation}
and for stimulated emission of an IR photon
\begin{equation}\label{eq9}
\begin{split}
\varphi_{s}^{(2-)}-\varphi_{d}^{(2-)}&=\varphi_{s}^{cc-}+\varphi^{bc}_{(>)}-\varphi_{d}^{cc-}-\varphi^{bc}_{(>)}\\
&=\varphi_{s}^{cc-}-\varphi_{d}^{cc-}.
\end{split}
\end{equation}
This enables to directly measure the influence of the final-state angular momentum on the cc-phase independent of the preceding one-photon bound-free transition.

 Indeed, in each case, we retrieve the phase difference  between pathways involving the same intermediate state, i.e., cc-transitions following the absorption of the same XUV photon. Consequently, the phase of the bc-transition, which includes both XUV chirp and the $p$-wave scattering phase, cancels out, such that the remaining phase difference is purely due to the one-photon transition in the continuum.
 
In contrast to the traditional RABBITT analysis \cite{Mairesse2003, Cattaneo2016a}, where phase differences $\varphi(E+\omega)-\varphi(E-\varphi)$  are extracted in order to approximate the phase derivative, our method yields an (absolute) phase difference at a fixed energy. In detail, in RABBITT, the total angle-integrated sideband phase contains the phase difference between pathways originating from neighbouring harmonics, i.e., an approximated phase derivative across two harmonics. Therefore, even when comparing different species, the measured delays correspond to differences of derivatives, or respectively, differences in group delay. As a consequence, absolute phase differences remain hidden. In contrast, by comparing pathways following the absorption of the same harmonic, the present procedure allows us to extract an absolute phase difference between two pathways.

Figure \ref{fig4} and \ref{fig5} show the mean of the experimentally retrieved phases for sidebands 18, 20 and 22, averaging over four independent measurements. The error bars indicate the uncertainty of the mean.
%

\section{Theoretical results}

\begin{figure}
\includegraphics[width=0.48\textwidth]{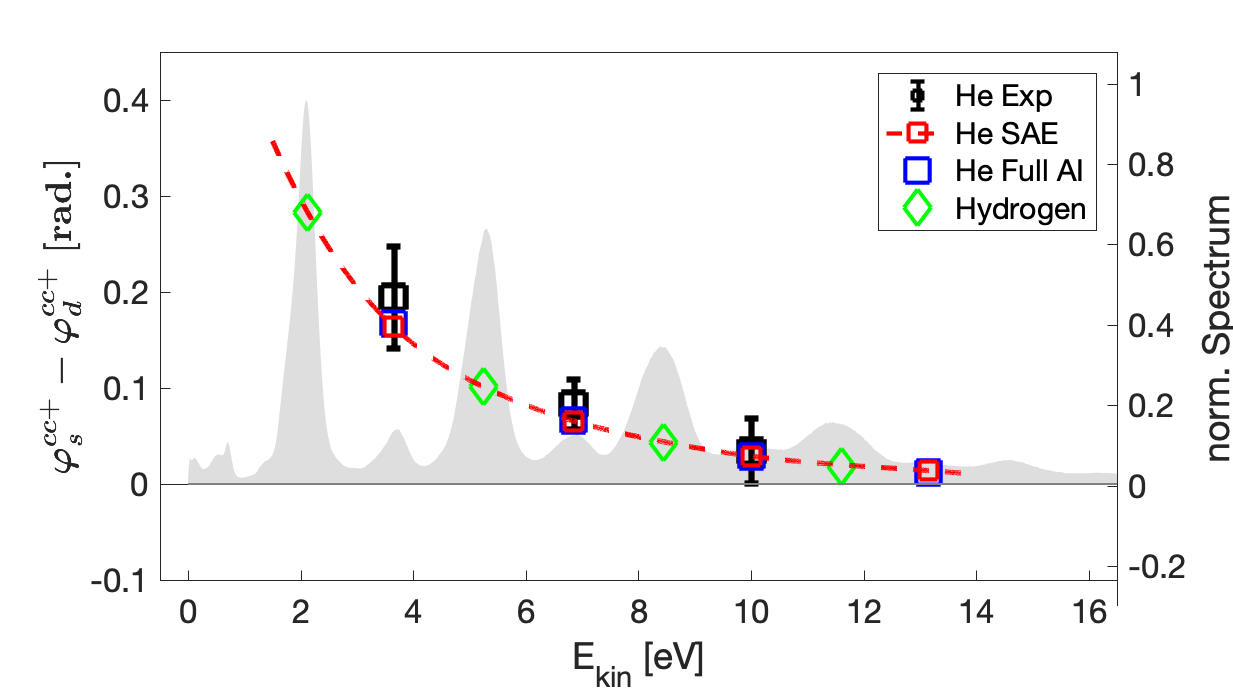}
\caption{\label{fig4} Difference  $\varphi_s^{cc+}-\varphi_d^{cc+}$ in radians for the cc-transition involving absorption. The experimental values represent the mean values of four measurements and the error bars correspond to the uncertainty of the mean. The discrepancy between the two simulations lies below 2\% of their absolute value. The shaded area represents the PES given in Fig. \ref{fig1}.}
\end{figure}

In order to prove the validity of the present extraction method, we apply the same fitting procedure to computed RABBITT traces, for which the total phase of the two-photon electron wave packet and thus the cc-phase can be directly accessed. We  performed Single Active Electron (SAE) calculations \cite{Grum-Grzhimailo2013,Douguet2016}, where the TDSE for helium is solved using the potential of Ref. \cite{Tong2005} and a finite-difference scheme based on the Crank-Nicolson method. The sideband signal is analysed in the same way as the experimental data and illustrated with the corresponding fit in Figure \ref{fig3} (center). The retrieved values for the phase difference between the $s$- and $d$-final-state partial waves are shown in Figures \ref{fig4} and \ref{fig5} for sidebands 18 to 24. In addition, by using a single harmonic in the TDSE calculations, we directly retrieve the phase of the outgoing $s$- and $d$-waves at the neighbouring sidebands without the need to invoke interferences of different pathways. Using the latter procedure, we find excellent agreement with the fitted phases, thereby confirming the validity of the present extraction method.

To probe for the possible influence of electron correlation and exclude possible shortcomings of the SAE approximation, we additionally perform full \textit{ab initio} simulations using the time-dependent close coupling method \cite{Pindzola2007} on a spatial FEDVR grid \cite{Rescigno2000} thereby solving the full two-electron TDSE for atomic helium from first principles \cite{Feist2008, Donsa2019}. The electric fields are treated in the dipole approximation. Both, the \textit{ab initio} and the SAE simulation employ an IR pulse with central wavelength of 790 nm and a Gaussian envelope with 8 fs FWHM. The spectral amplitude and phase of each harmonic was chosen to match the experimental spectrum. We have checked for the potential influence of the IR pulse duration on the extracted phases. The excellent agreement of the results from SAE simulations for two different IR pulse durations of 8 fs and 20 fs (FWHM) allows one to rule out any significant pulse duration effects on the resulting phases.

Comparing the results of the two independent simulation methods with the experiment in Figs. \ref{fig4} and \ref{fig5}, we observe excellent agreement between all three data sets. We therefore can conclude that the effect of electron correlation on the cc-transition is negligible or, respectively, identical for the compared pathways, in the investigated energy range.

In addition, we report calculations for the hydrogen atom, for which harmonics from the 9$^{th}$ to the 17$^{th}$ order are used to generate the XUV spectrum, such that the electron kinetic energy remains in the same range as for helium. It can be observed in Figures \ref{fig4} and \ref{fig5}, that the retrieved phase delays exactly follow the helium trend, thus supporting the argument of negligible influence from both electron correlation effects and the helium short-range potential on the investigated cc-transition time delays.

\section{Discussion}

We observe a remarkable quantitative agreement between the experiment and theoretical values for the phase difference $\varphi_s^{cc\pm}-\varphi_d^{cc\pm}$ for electron energies 2 eV$\leq$E$\leq$14 eV obtained from two independent computational methods for both hydrogen and helium. We have found a significant phase difference with a maximum value of 0.21 rad between s-and d-partial waves at harmonic 18 corresponding to a final-state with energy of 3.7 eV. Our data reveals three main features:

\renewcommand{\labelenumi}{\roman{enumi}.}
\begin{enumerate}
\item The relative phase between the s- and d partial-wave is ubiquitous positive and decreases with energy, both for cc-transitions involving absorption and stimulated emission at all kinetic energies.
\item At all sidebands, the absolute values of the phase difference between $s$- and $d$-wave are almost equal for absorption and stimulated emission. The discrepancy lies far below the experimentally accessible precision. The theoretical values indicate slightly larger delays for absorption.
\item	The cc-phase difference converges to zero with increasing kinetic energy.
\end{enumerate}

\begin{figure}
\includegraphics[width=0.48\textwidth]{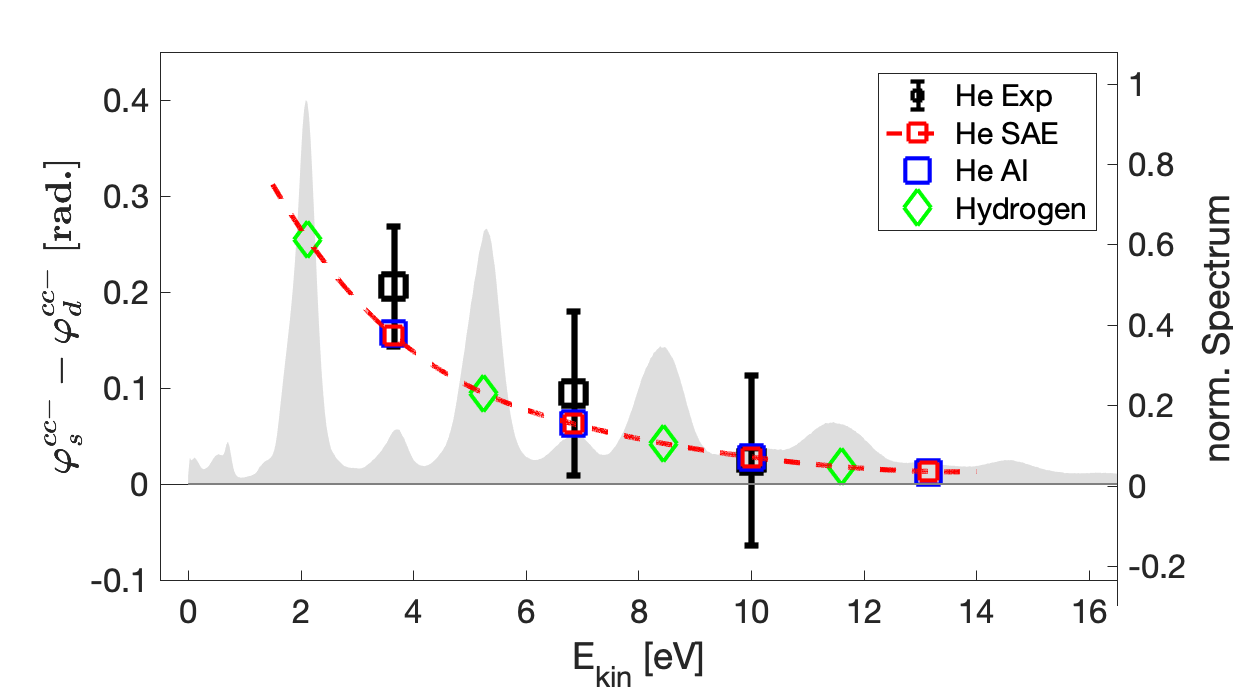}
\caption{\label{fig5} Difference $\varphi_s^{cc-}-\varphi_d^{cc-}$ for the cc-transition involving stimulated emission, otherwise same as Figure \ref{fig4}.}
\end{figure}
	
Although the observed phase difference has previously not been recognized and analyzed, the above observations are, in fact, noticeable also in earlier numerical simulations \cite{Dahlstrom2012,Ivanov2017}. Qualitatively the trend mirrors the observation of partial-wave interferences resulting in electron-spin polarization in one-photon bc-transitions \cite{Heinzmann2012, Heinzmann2013, Fanciulli2017}.
The observations (i)-(iii) are fully consistent with the influence of the final-state centrifugal potential on the continuum scattering phase and on the EWS time delay, in the present case for continuum-continuum transitions. Observation (iii) is the obvious consequence of the decreasing effect of the underlying potential energy landscape on the escaping electron. For increasing energies, the wave function tends towards the behaviour of a free spherical wave for which all delays vanish. Moreover, with increasing momentum of the outgoing wavepacket, the cc-transition is effectively shifted to larger distances from the ionic core at which the centrifugal potential $\propto 1/r^2$ becomes negligible compared to the Coulomb potential. The latter was the underpinning of the previous analytic estimates of the cc-phase and time delay in which the angular momentum dependence was neglected \cite{Klunder2011}.

Observation (i) clearly shows that the cc-phase is in fact, directly related to the EWS phase for cc-scattering. This is supported by the observation that the phase and corresponding delay qualitatively resembles the EWS delay for bound-continuum transitions to different angular momentum states \cite{Heinzmann2012, Heinzmann2013, Fanciulli2017}. The fact that the retrieved phases are significantly smaller (by factors 3 to 4) with respect to the scattering phase is due to the fact that, unlike for the bc-transition, the cc-transition in the two-photon scenario probes the potential landscape not for the full half-scattering but only at large distances where the centrifugal potential is weaker, yet still leads to clearly resolvable effects at low energies.

Observation (ii) then confirms the fact that the relevant phase is accumulated at distances where the Coulomb potential $(1/r)$ dominates and the centrifugal potential provides a (small) correction, short-ranged contributions are entirely negligible. Therefore, the observed phases are universal, i.e. independent of the atomic species, and slightly larger for absorption than for emission. The latter is in line with the fact that the outgoing wavepacket after the bc-transition propagates initially slower before absorption thereby enhancing the influence of the centrifugal potential on the subsequent cc-transition.

The resulting EWS delay between the $s$- and $d$-partial wave, observed for the first time in a cc-transition, allows for a simple, quasi-classical interpretation: Due to the different angular momentum the rotational and radial energy distribution of the $s$- and $d$-wave packet components are different, and, since the rotational energy fraction is larger for the $d$-wave components, the radial expansion is slower. This implies a positive EWS delay $\frac{d}{dE} (\delta_{l=2} (E)-\delta_{l=0} (E))$ of the $d$-wave relative to the $s$-wave, consistent with well-known trends for EWS delays in bound-continuum transitions. More generally, one expects larger delays for wave packet components with higher angular quantum numbers. The quantitative experimental confirmation of this effect for cc-transitions is the main finding of this work.
\begin{figure}
\includegraphics[width=0.48\textwidth]{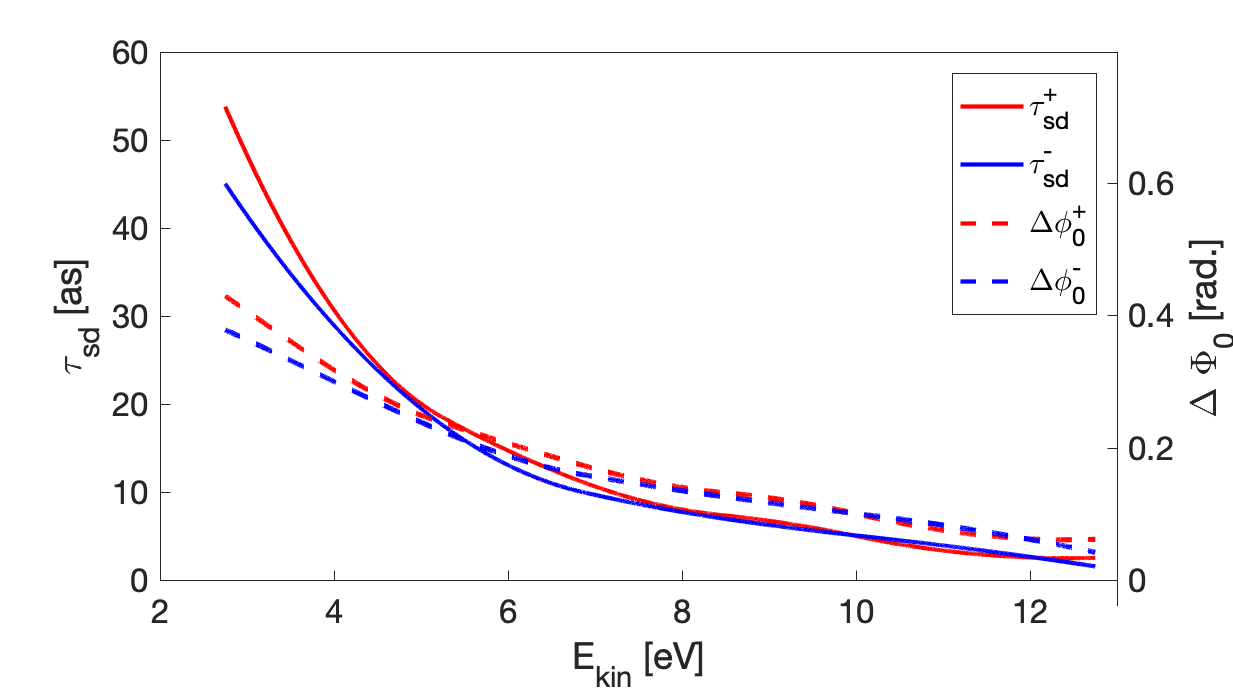}
\caption{\label{fig6} Delay $\tau_{sd}$ and absolute phase difference $\Delta\phi_0$ between the outgoing d- and s-partial waves as a function of the wave packets center energy.}
\end{figure}
For an electron wave packet centered at 6.1 eV, as in the present experiment, the $d$-wave is retarded by 12 as with respect to the $s$-wave. Higher order derivatives of the measured phase difference affect the shape of the envelope and lead to differences in chirp between the $s$- and $d$-partial waves. The implication for the wave packet in time can be inferred from the Fourier integral over the spectral components of absorption and emission pathways of the two partial wave packets. Using the energy-dependent phase difference $\Delta \varphi_{sd}$ the partial wave functions can be written in time domain as
\begin{equation}\label{eq11}
\phi_s(t)=\int A_s^{(2)}(\omega)e^{-i\omega t}d\omega
\end{equation}
and
\begin{equation}\label{eq12}
\phi_d(t)=\int c_{d/s}(\omega)A_s^{(2)}(\omega)e^{-i\Delta \varphi_{sd}(\omega)}e^{-i\omega t}d\omega
\end{equation}
where $\omega=E_{kin}/\hbar$ and $c_{d/s}=|A_d^{(2)}|/|A_s^{(2)}|$ is the absolute amplitude ratio. Approximating the phase difference with a first order Taylor expansion $\Delta\varphi_{sd} (\omega)=\Delta\varphi_{sd} (\omega_0 )+\Delta \varphi_{sd}^{\prime} |_{\omega_0} (\omega-\omega_0)$ around the center (mean) frequency $\omega_0$ of the wave packet and assuming a weakly energy dependent amplitude ratio $c_{d/s}$ it follows immediately that
\begin{equation}\label{eq13}
\begin{split}
\phi_d(t)&= c_{d/s}(\omega_0)e^{i[\Delta \varphi_{sd}^{\prime}|_{\omega_0}\omega_0-\Delta \varphi_{sd}(\omega_0)]}\\
&\hspace{3cm} \int A_s^{(2)}(\omega)e^{-i\omega(t+\Delta \varphi_{sd}^{\prime}|_{\omega_0})}d\omega \\[10pt]
&=\phi_s(t+\Delta \varphi_{sd}^{\prime}|_{\omega_0})c_{d/s}(\omega_0)e^{-i\Delta\phi_{0}}
\end{split}
\end{equation}
where $\Delta \phi_0= \Delta \varphi_{sd} (\omega_0)-\Delta \varphi_{sd}^{\prime} |_{\omega_0} \omega_0$. The assumption of a flat amplitude ratio is crucial and only valid for a narrow energy spectrum. An explicit study of the amplitude ratio for various species is given in \cite{Haber2011, Busto2018}).

Hence, from Eq. \eqref{eq13} it follows that, additionally to the group delay $\tau_{sd}=-\Delta\varphi_{sd}^{\prime} |_{\omega_0}$, an additional offset phase $\Delta\phi_0$ is imparted to $d$- with respect to the $s$-partial wave. This additional phase shift corresponds to an advance of the wave packets absolute phase with respect to the wave packet envelope. This phase lag between the s and the d wave in the outgoing wave packet implies an interesting analogue to the carrier-envelope phase (CEP) slip. Although the absolute phase has no impact for the electrons classical observables, i.e. localisation and momentum, it determines the interference. To the best of our knowledge we hereby report for the first time an effect of an electron wave packets absolute phase. It is illustrated together with the group delay $\tau_{sd}$ as function of the center energy in figure 6.

As can be inferred from Figures \ref{fig4} and \ref{fig5}, the theoretically calculated cc-phase difference in hydrogen follows the same trend as in helium. This finding confirms the fact that the relative phase between wave packet components with different angular momenta is a universal property. For larger atoms or molecules, where electron correlation effects become dominant, a deviation from the observed trend cannot be excluded though.

At lower kinetic energies, even larger delays are to be expected. These were not measured in this work due to the limited tunability of the XUV spectrum in the present experiment. Sideband 16, in principle, would lie just above the helium ionisation threshold and could be analysed along those lines. However, higher excited states of the neutral helium atom come into play here \cite{Lucchini2015} noticeable in Figure \ref{fig2} (bottom). Including the latter sideband would then involve more complex transitions  beyond cc-transitions and is beyond the scope of this work. 

\section{Conclusion and outlook}

In conclusion, we have established an experimental protocol which allows us to disentangle the contributions from up to four different photoionisation pathways in atomic helium leading to the same final energy.  With a novel fitting technique which uses both the phase and the amplitude of the angular resolved RABBITT interference pattern, we have been able to determine the amplitudes and the relative phases of all four quantum pathways that contribute to each sideband. Comparing pathways following the absorption of the same XUV photon, we find a time delay between $s$- and $d$-waves arising from one-photon transitions in the continuum as large as 12 attoseconds. This represents the first measurement of the EWS time delay for (inverse) Bremsstrahlung. Moreover, we find excellent quantitative agreement between  the experiment and two independent theoretical simulations. The observed trend reveals ubiquitous positive phase delays between $s$- and $d$-waves for both absorption and stimulated emission. The measured relative phase, which vanishes for high kinetic energies, is determined by the final state of the continuum wave packet components with different angular momentum populated by the two-photon transition. The radiative transition in the continuum occurs at large distances where the Coulomb potential of the nearby ion and the centrifugal potential dominate over the target-dependent short-range potential. As a consequence, the relative phases are expected to be a universal property of radiative transitions in the continuum that is relevant to characterize the photoemission dynamics for different atomic species. The same absolute phase difference affects the sideband anisotropy even in the stationary regime, as for example in the laser-assisted ionization of helium with monochromatic synchrotron radiation.

This work not only serves as a proof-of-principle demonstration for accurately disentangling multiple interfering quantum pathways but also gives new physical insight into the time properties of the fundamental inverse and stimulated Bremsstrahlungs process. The proposed method can be easily generalised to other systems and cc-transitions. The work opens up new experimental opportunities for analysing and selecting quantum pathways in larger systems such as heavier atoms and molecules, where different quantum pathways can lead to distinct molecular breakup reactions or final states. Additionally, we hope that our study will motivate further  experimental and theoretical studies of cc-transitions not only in various atomic species, but more generally in small molecules, aiming for a general understanding of intermediate to long-range interactions on the photoemission time delay.

\begin{acknowledgments}
J.F., L.C. and U.K. acknowledge the support of the NCCR MUST, funded by the Swiss National Science Foundation.
L.A. and N.D. acknowledge the support of the United States National Science Foundation under NSF Grant No. PHY-1607588 and by the UCF in-house OR Grant Acc. No. 24089045.
Parts of the calculations were performed on the Vienna Scientific Cluster (VSC3). S.D. and J.B. acknowledge the support by the WWTF through Project No. MA14-002, and the FWF through Projects No. FWF-SFB041-VICOM, and No. FWF-W1243-Solids4Fun, as well as the IMPRS-APS. 
F.M. acknowledges the MINECO projects FIS2016-77889-R, the 'Severo Ochoa' Programme for Centres of Excellence in R{\&}D (SEV-2016-0686) and the 'María de Maeztu' Programme for Units of Excellence in R{\&}D (MDM-2014-0377).
\end{acknowledgments}

\bibliography{arxiv_manuscript}

\end{document}